\documentclass[10pt]{ismd08}
\usepackage{graphicx}
\usepackage{cite,mcite}
\usepackage{xspace}
\usepackage{hepnames,units}
\usepackage{floatflt}

\newcommand{\atlas}{ATLAS\xspace}
\newcommand{\thetastar}{\ensuremath{\theta^{*}}\xspace}
\newcommand{\thetastarll}{\ensuremath{\theta^{*}_{ll}}\xspace}

\newcommand{\fig}{Fig.}
\newcommand{\tab}{Tab.}

\newcommand{\msugra}{mSUGRA\xspace}

\newcommand{\ET}{\ensuremath{E_T}\xspace}
\newcommand{\ETmis}{\ensuremath{E_T^{\mathrm{mis}}}} % xxx
\newcommand{\Meff}{\ensuremath{m_\mathrm{eff}}} 
\newcommand{\Mhalf}{\ensuremath{m_{1/2}}} \newcommand{\mhalf}{\Mhalf}
\newcommand{\Mzero}{\ensuremath{m_{0}}} \newcommand{\mzero}{\Mzero}
\newcommand{\mll}{\ensuremath{m_{\Plepton\Plepton}}\xspace}
\newcommand{\mtautau}{\ensuremath{m_{\Ptau\Ptau}}\xspace}
\newcommand{\mllmax}{\ensuremath{\mll^{\mathrm{max}}}}
\newcommand{\mql}[1]{\ensuremath{m_{\Pquark\Plepton^\mathrm{#1}}}}
\newcommand{\mqbarl}{\ensuremath{m_{\APquark\Pleptonmp}}\xspace}
\newcommand{\mqlmax}[1]{\ensuremath{m_{\Pquark\Plepton\mathrm{#1}}^{\mathrm{max}}}}
\newcommand{\mqll}[1]{\ensuremath{m_{\Pquark\Plepton\Plepton}^{\mathrm{#1}}}}
\newcommand{\mqllmax}{\mqll{max}}
\newcommand{\mqllmin}{\mqll{min}}

\newcommand{\invfb}{fb\ensuremath{^{-1}}}

\begin{document}
\title{Supersymmetry and other beyond the Standard Model physics:
  Prospects for determining mass, spin and CP properties}
\author{Wolfgang Ehrenfeld$^1$ for the \atlas Collaboration}
\institute{$^1$Deutsches Elektronen Synchrotron, Notkestrasse 85,
  22603 Hamburg, Germany}
\maketitle
\begin{abstract}
  The prospects of measuring masses, spin and CP properties within
  Supersymmetry and other beyond the Standard Model extensions at the
  LHC are reviewed. Emphasis is put on models with missing transverse
  energy due to undetected particles, as in Supersymmetry or Universal
  Extra Dimensions.
\end{abstract}

\section{Introduction}

It is widely expected that the Large Hadron Collider~(LHC), which
started very successfully on the 10th September 2008 with single
beam injection, will uncover physics beyond the present Standard Model~(SM) of
particle physics. Supersymmetry~(SUSY) is one of the most promising
candidates for new physics. Among its virtues are the potential to
overcome the hierarchy problem, to provide a dark matter candidate and
make a unification of gauge coupling constants at a high energy scale
possible. If the SUSY mass scale is in the sub-TeV range, already
first LHC data will likely be sufficient to claim a discovery of new physics
although new physics do not strictly mean SUSY as 
other new physics scenarios can have similar features and properties.
In order to distinguish different scenarios of new physics
and to determine the full set of model parameters within one scenario 
as many measurements of the new observed phenomena as possible are
needed. This includes the precise measurement of masses, spins and CP
properties of the newly observed particles.  

Both multi-purpose experiments at the LHC, \atlas\cite{det:atlas} and
CMS\cite{det:cms}, are designed for these measurements. They will be
able to pin down the exact model of new physics, e.~g. to distinguish
SUSY from Universal Extra Dimensions~(UED).

\section{Supersymmetry}

In the following we assume R-parity conservation. As a consequence
sparticles can only be produced in pairs and the lightest SUSY
particle~(LSP) is stable, which usually escapes detection in
high-energy physics detectors. At LHC energies mostly pairs of squarks or
gluinos are produced in proton-proton collisions, which then
subsequently decay via long cascades into the LSP. Typical event
topologies at the LHC are multi jet events with zero or more leptons
and missing transverse energy due to the two LSPs. 
In the case of \atlas these events will be triggered using a combined
jet and missing \ET trigger. The selection is mainly based on four
jets ($p_T^{j_1} > \unit[100]{GeV}$, $p_T^{j_2,j_3,j_4} >
\unit[50]{GeV}$) and missing \ET ($\ETmis > \unit[100]{GeV}, 
0.2\Meff$). The effective mass, \Meff, is the scalar sum of missing
\ET and the transverse momentum of the four leading
jets. For further details see~\cite{atlas:csc}. With this kind of
selection minimal Supergravity (\msugra) models up to $\Mhalf \sim
\unit[0.7]{TeV}$ or $\Mzero \sim \unit[3]{TeV}$ can be discovered with
a luminosity of \unit[1]{\invfb}.

\subsection{Mass Measurements}
\label{sec:mass-mearuement}

\begin{floatingfigure}{0.45\textwidth}
  \centering
  \unitlength=10mm
  \vspace*{-7mm}
  \begin{picture}(4,2)
    \put(0.0,0.75){\vector(4,1){1}}
    \put(0.25,0.25){$\Psquark_{L}$}
    \put(1.0,1.00){\vector(1,0){1}}
    \put(1.0,1.00){\vector(1,4){0.25}}
    \put(0.75,1.75){\Pq}
    \put(1.25,0.50){\PSneutralinoTwo}
    \put(2.0,1.00){\vector(4,-1){1}}
    \put(2.0,1.00){\vector(1,2){0.5}}
    \put(2.0,1.75){\Pleptonplus}
    \put(2.25,0.25){$\Pslepton^-_{R}$}
    \put(3.0,0.75){\vector(2,-1){1}}
    \put(3.0,0.75){\vector(1,1){.75}}
    \put(3.25,1.5){\Pleptonminus}
    \put(3.25,0.){\PSneutralinoOne}
  \end{picture}
  \vspace*{-2mm}
  \caption{Prime example of a SUSY decay chain for SUSY mass
    reconstruction. The first lepton in the decay chain is called the \emph{near}
    lepton while the other is called the \emph{far} lepton.}
  \label{fig:quarkleptonlepton}
\end{floatingfigure}
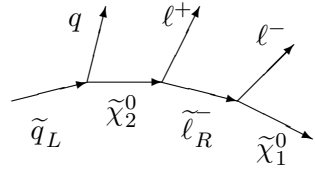
After the discovery of new physics beyond the SM as many measurements
of the production process and particle properties are needed to
pin-down the exact model of new physics. For example the masses of the new
particles can be used to distinguish between different SUSY models.
Due to the two escaping LSPs in every SUSY event, no mass peaks can be
reconstructed and masses must be measured by other means.
In \msugra models the main source of mass information is provided by
\PSneutralinoTwo decays, such as $\PSneutralinoTwo \to
\PSlepton^\pm\Plepton^\mp \to \PSneutralinoOne\Plepton^+\Plepton^-$
(see \fig~\ref{fig:quarkleptonlepton}).
First we consider the invariant mass spectrum of the two leptons
$\mll$ from the decay chain in
\fig~\ref{fig:quarkleptonlepton}. Due to the scalar nature of the
slepton, the invariant mass exhibits a triangular shape with a sharp
drop-off at a maximal value $\mllmax$.  The position of this endpoint
depends on the masses of the involved sparticles:
\begin{equation}
  \label{eq:3}
  \mllmax = m_{\PSneutralinoTwo} 
            \sqrt{1-\left(\frac{m_{\PSlepton_R}}{m_{\PSneutralinoTwo}}\right)^2}
            \sqrt{1-\left(\frac{m_{\PSneutralinoOne}}{m_{\PSlepton_R}}\right)^2}.
\end{equation}
Combinatorial background from SM and other SUSY processes is
subtracted using the flavor-subtraction method. The endpoint is
measured from the di-lepton (electron and muon) mass distribution
$N(\Pelectron\Ppositron)/\beta + \beta N(\Pmuon\APmuon) -
N(\Pepm\Pmump)$, where $N$ is the number of selected events and
$\beta$ is the ratio of the electron and muon reconstruction
efficiency ($\beta\simeq 0.86$)~\cite{atlas:csc}.
Figure~\ref{fig:invmass:ll} shows the mass distribution for different
\msugra benchmark points\footnote{Within \atlas the \msugra benchmark
  points are called SU\emph{X}. 
  SU1: $\mzero=\unit[70]{GeV},\mhalf=\unit[350]{GeV},A_0=0,\tan\beta=10,\mu>0$
  SU3: $\mzero=\unit[100]{GeV},\mhalf=\unit[300]{GeV},A_0=-300,\tan\beta=6,\mu>0$
  SU4: $\mzero=\unit[200]{GeV},\mhalf=\unit[160]{GeV},A_0=-400,\tan\beta=10,\mu>0$
}. 
The SU3 point is an
example of a simple two-body decay (\fig~\ref{fig:invmass:ll}(b)), SU4 illustrates a more complex
three-body decay (\fig~\ref{fig:invmass:ll}(c)) and SU1 two two-body
decays (\fig~\ref{fig:invmass:ll}(a)). In all cases the \mll
endpoint can be measured without a bias although the needed luminosity
is quite different. Further, the fit function to extract the
endpoint(s) needs to be adjusted to the underlying mass spectrum. The expected sensitivity is
summarized in \tab~\ref{tab:masses} including the assumed luminosity.
\begin{figure}[ht]
  \centering
  \includegraphics[width=0.45\textwidth]{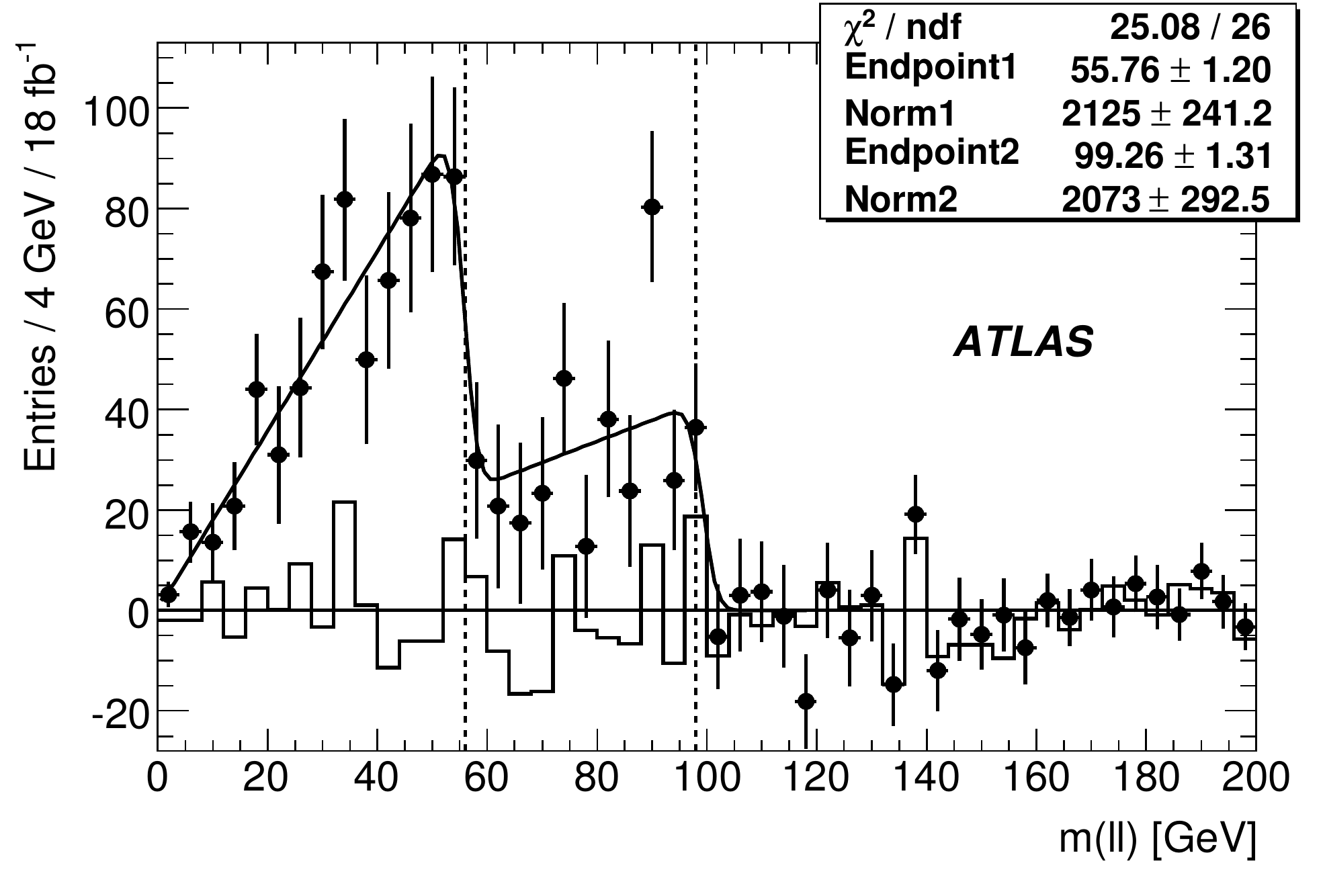}\hfill%
  \includegraphics[width=0.45\textwidth]{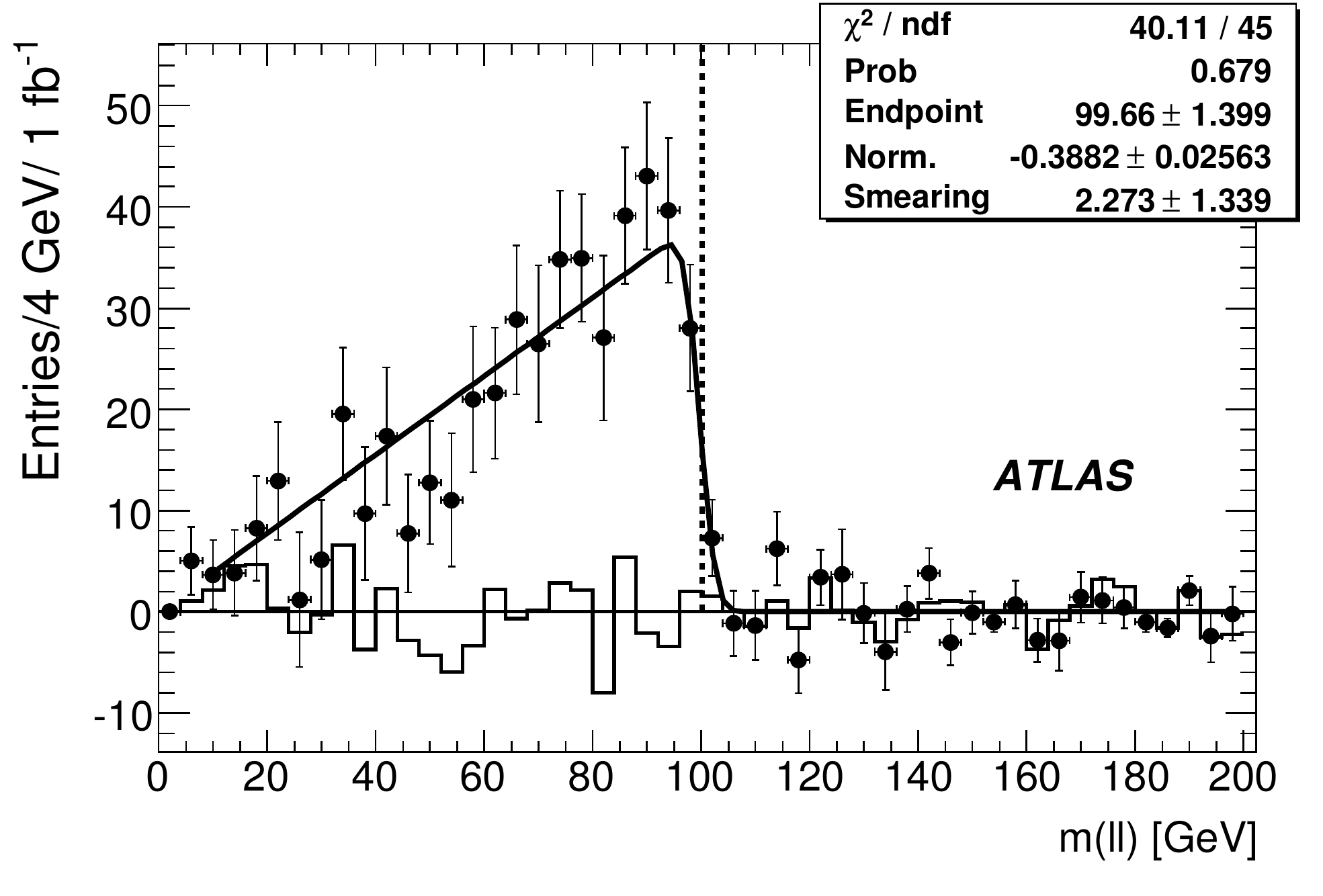}\\
  \includegraphics[width=0.45\textwidth]{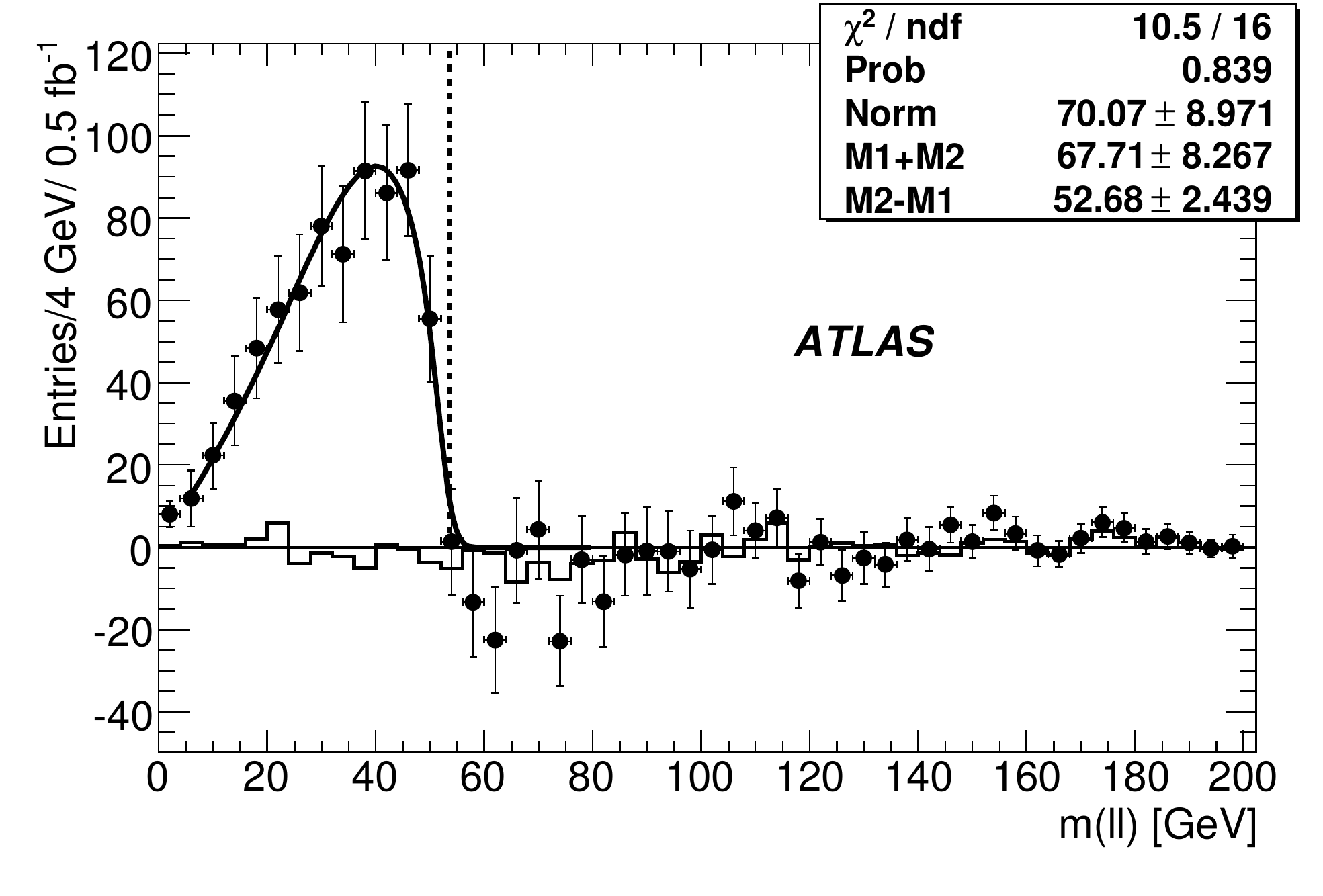}\hfill%
  \includegraphics[width=0.45\textwidth]{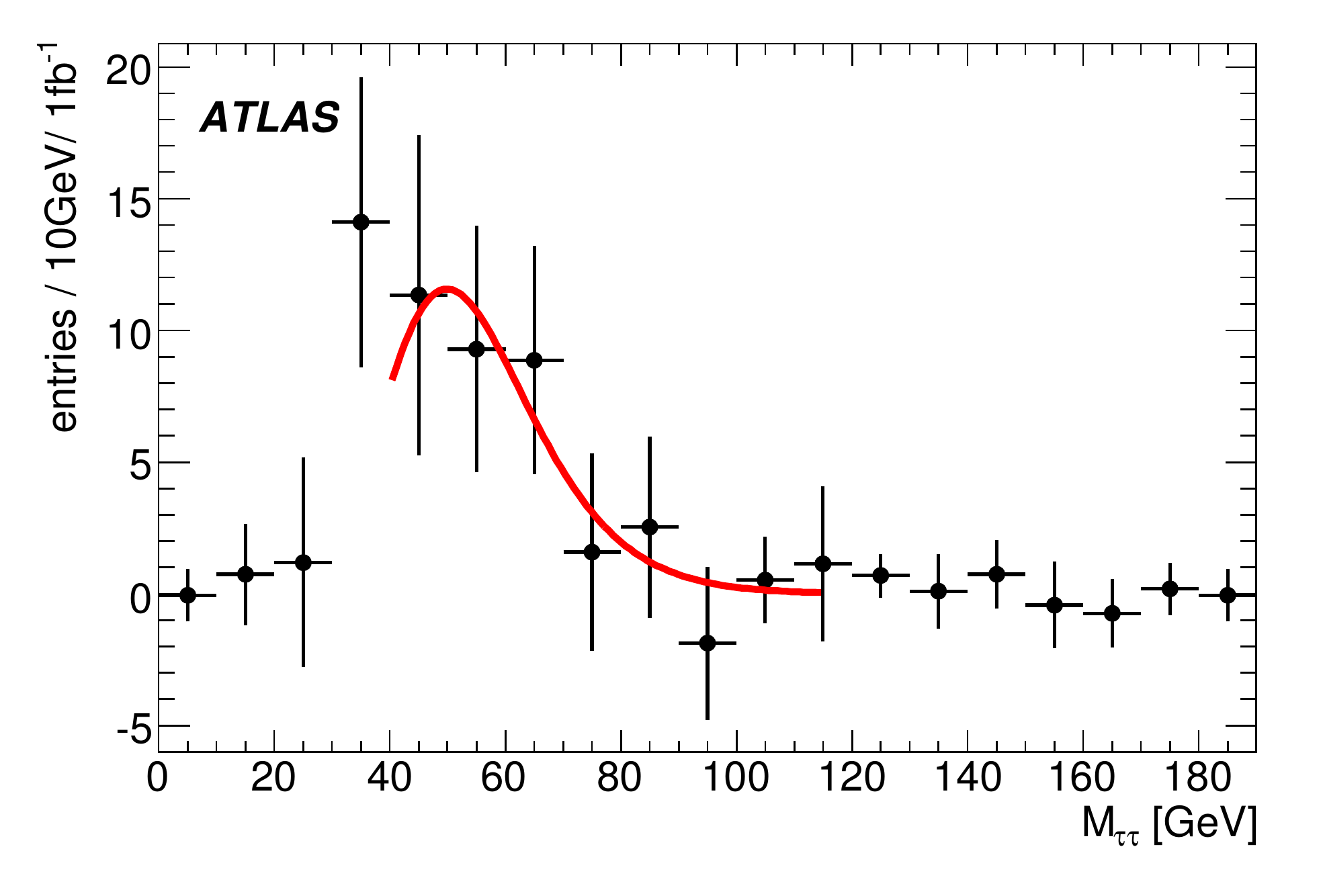}\\[-7mm]
  \begin{picture}(0,0)
    \put(-178,230){\textsf{\textbf{\footnotesize(a)}}}
    \put(55,230){\textsf{\textbf{\footnotesize(b)}}}
    \put(-178,25){\textsf{\textbf{\footnotesize(c)}}}
    \put(55,25){\textsf{\textbf{\footnotesize(d)}}}
  \end{picture}
  \caption{Flavour subtracted di-lepton mass spectrum for different
    \msugra benchmark points: (a) SU1($\Plepton=e,\Pmu$), (b)
    SU3($\Plepton=e,\Pmu$), (c) SU4($\Plepton=e,\Pmu$), (d)
    SU3($\Plepton=\Ptau$).}
  \label{fig:invmass:ll}
\end{figure}
\begin{table}[ht]
  \footnotesize
  \centering
  \begin{tabular}{llrrr}
    observable & benchmark point & \multicolumn{1}{l}{true mass \unit{[GeV]}} & \multicolumn{1}{l}{expected mass \unit{[GeV]}} & \multicolumn{1}{l}{luminosity \unit{[\invfb]}}\\
    \hline
    \mll & SU1     & 56.1  & $55.8 \pm 1.2 \pm 0.2$     & 18 \\
    \mll & SU1     & 97.9  & $99.3 \pm 1.3 \pm 0.3$     & 18 \\
    \mll & SU3     & 100.2 & $99.7 \pm 1.4 \pm 0.3$     & 1.0 \\
    \mtautau & SU3 & 98    & $102 \pm 17 \pm 5.5 \pm 7$ & 1.0 \\
    \mll & SU4     & 53.6  & $52.7 \pm 2.4 \pm 0.2$     & 0.5 \\
    \hline
    \mqlmax{(low)}  & SU3 & 325 & $333 \pm 6 \pm 6 \pm 8$    & 1.0 \\
    \mqlmax{(high)} & SU3 & 418 & $445 \pm 11 \pm 11 \pm 11$ & 1.0 \\
    \mqllmin      & SU3 & 249 & $265 \pm 17 \pm 15 \pm 7$  & 1.0 \\
    \mqllmax      & SU3 & 501 & $501 \pm 30 \pm 10 \pm 13$ & 1.0 \\    
    \hline
  \end{tabular}
  \caption{Reconstructed endpoint positions. The first error of the
    expected value is the statistical error and the second is due to
    systematic from  the lepton energy scale and $\beta$. In case of
    \mtautau the third error is due to the uncertainty in the tau polarization.}
  \label{tab:masses}
\end{table}
A similar analysis can be performed if we replaced electrons and muons
by taus. Due to the additional neutrinos from the tau decay, the
visible di-tau mass distribution is not triangular any more (see
\fig~\ref{fig:invmass:ll}(d)). This complicates measuring the endpoint of
the spectrum. A solution to this problem is to fit a suitable function
to the trailing edge of the visible di-tau mass spectrum and use the
inflection point as an endpoint sensitive observable, which can be
related to the true endpoint using a simple MC based calibration
procedure. Figure~\ref{fig:invmass:ll}(d) shows the charge
subtracted  visible di-tau mass distribution $N(\Ptauon\APtauon) -
N(\Ptaupm\Ptaupm)$ which is used to suppress background from fake taus
and combinatorial background. The expected sensitivity is listed in
\tab~\ref{tab:masses}. Please note, that the third error is due to the
SUSY-model dependent polarization of the two taus. On the other hand
this influence of the tau polarization on the di-tau mass distribution
can be used to measure the tau polarization from the mass distribution
and distinguish different SUSY models from each other.

By including the jet produced in association with the \PSneutralinoTwo
in the $\PSq_L$ decay (see \fig~\ref{fig:quarkleptonlepton}),
several other endpoints of measurable mass combinations are possible:
\mqlmax{(low)}, \mqlmax{(high)}, \mqllmin, \mqllmax. The label min/max
denotes the upper/lower endpoint of the spectrum. In the case of
\mqlmax{} the near and the far lepton can not be distinguished in most
of the SUSY models and instead the minimum/maximum of the mass
$m_{\Pquark\Pleptonpm}$ is used. As in the di-lepton case a suitable
fit function for each observable is needed. The expected sensitivity to the different
mass combinations for the SU3 model are summarized in
\tab~\ref{tab:masses}.

These five mass combinations can be used to extract the underlying high
mass model parameters using fitting programs like Fittino\cite{Bechtle:2004pc}
or SFitter\cite{Lafaye:2007vs}.

\subsection{Spin Measurements}
\label{sec:spin-measurements}

Measuring the number of new particles and their masses will give us
enough information to extract model parameters for a certain extension
of the SM. However, the mass information will not always be enough to
distinguish different scenarios of new physics. For example, UED with
Kaluza-Klein (KK) parity can be tuned in such a way that it reproduces
the mass spectrum of certain SUSY models. However, the spin of the new
particles is different and can be used to discriminate between these
models.

\begin{figure}[b]
\begin{minipage}[t]{0.58\linewidth}
  \centering
  \includegraphics[width=1.78\textwidth,page=19,viewport=320 595 800 760,clip]{phys-pub-2007-004.pdf}
  \begin{picture}(0,0)
    \put(0,50){\textsf{\textit{\textbf{\footnotesize\atlas}}}}
  \end{picture}
  \vspace*{-5mm}
  \caption{Expected charge asymmetry $A$ for SU3 and 
    \unit[30]{\invfb}.}
  \label{fig:chargeA}
\end{minipage}\hfill%
\begin{minipage}[t]{0.4\linewidth}
  \centering
  \includegraphics[width=0.88\textwidth]{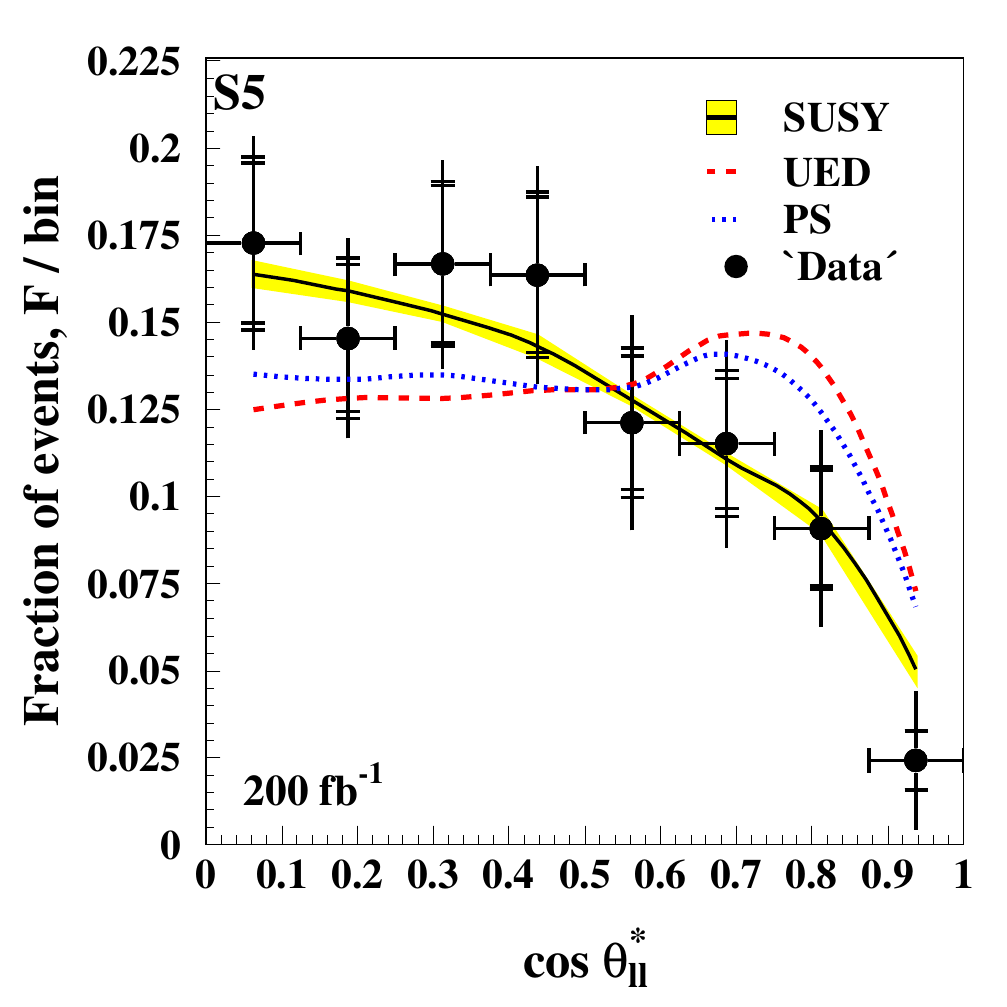}
  \begin{picture}(0,0)
    \put(-80,35){\textsf{\textit{\textbf{\footnotesize\atlas}}}}
  \end{picture}
  \vspace*{-5mm}
  \caption{Expected \thetastarll distribution for the S5 bench mark
    point and \unit[200]{\invfb}.}
  \label{fig:thetastarll}
\end{minipage}
\end{figure}

The standard SUSY decay chain (see \fig~\ref{fig:quarkleptonlepton})
can also be used to measure the spin of
\PSneutralinoTwo\cite{Barr:2004ze}. A charge
asymmetry $A$ is expected in the invariant masses \mql{near(\pm)}
formed by the quark and the near lepton. It is defined as $A=(s^+
-s^-)/(s^+ +s^-)$, where $s^{\pm} = d\sigma/d\mql{near(\pm)}$.  In
most of the cases it is experimentally not possible to distinguish
between near and far lepton and hence only $\mql{\pm}$ can be
measured, diluting $A$. Further, the
asymmetry from the corresponding \mqbarl charge distribution is the
same as the asymmetry for \mql{\pm}, but with opposite sign. Usually it is
not possible to distinguish \Pquark jets from \APquark jets at the
LHC. On the other side more squarks than anti-squarks will be produced.
The expected asymmetry $A$ for SU3
is shown in \fig~\ref{fig:chargeA} for a luminosity of
\unit[30]{\invfb}, where already \unit[10]{\invfb} are sufficient to
exclude the zero spin hypothesis at 99\% CL\cite{Biglietti:2007mj}. In the case of SU1
far and near leptons are distinguishable on
kinematic grounds. On the other hand, cross section times branching ratio of
this decay chain is much lower than the SU3 case, so that \unit[100]{\invfb}
are needed to exclude the zero spin hypothesis at 99\% CL.

\enlargethispage{2mm}%
The slepton spin can be measured in direct di-slepton production
$\Pquark\APquark \to \PZ\Pphoton \to \PSlepton\PSlepton \to
\PSneutralino\Plepton\PSneutralino\Plepton$. In UED the corresponding
process is $\Pquark\APquark \to \PZ\Pphoton \to \Plepton_1\Plepton_1
\to \Pphoton_1\Plepton\Pphoton_1\Plepton$, where $\Plepton_1$ and
$\Pphoton_1$ are the KK-lepton and -photon, respectively. Both have
spin $1/2$, the same as their SM partners. In both decay chains a SM
lepton-pair is produced, all other particles escape undetected.
Although the involved new particle masses can be the same, the slepton
spin ($0$) and KK-lepton spin ($1/2$) are different.
The angle \thetastar, as defined between the incoming quark and the
slepton/KK-lepton, can be used to discriminate between both model. The pure
phase space (PS) distribution would be flat. In SUSY and UED models it is
proportional to $1+A\cos^2\thetastar$, where $A=-1$ for SUSY and
$A=(E^2_{\Plepton_1}-m^2_{\Plepton_1})/(E^2_{\Plepton_1}+m^2_{\Plepton_1})$
for UED.
However, \thetastar is not directly accessible. Experimentally only
$\thetastarll \equiv \cos\left(2\tan^{-1} \exp
  \left(\Delta\eta_{\Pleptonplus\Pleptonminus}/2\right)\right) =
  \tanh\left(\Delta\eta_{\Pleptonplus\Pleptonminus}/2\right) 
$, the angle between the two leptons, can be measured. Note,
that \thetastarll is invariant under boosts along the beam axis. Still,
\thetastarll has some correlation with \thetastar.
Events with two good leptons ($p_T^{l_1,l_2}>\unit[40,30]{GeV}$) and
missing $\ET > \unit[100]{GeV}$ are selected. Further, events with
b-jets and high $p_T$ jets ($p_T>\unit[100]{GeV}$) are rejected\cite{Barr:2005dz}.
The expected \thetastarll distribution for a luminosity of
\unit[200]{\invfb} is shown in \fig~\ref{fig:thetastarll} including
the predictions for the SUSY, UED and PS case. Clearly, the difference
between all three cases can be see. For a five sigma significance
\unit[200]{\invfb} are needed to distinguish between SUSY and UED and
\unit[350]{\invfb} to distinguish between SUSY and PS.

\section{Other Beyond the Standard Model Physics}

The previous section was devoted to the measurement of masses and
spins in the case of missing energy due to non detectable new
particles within cascade decays. Without this complication the
measurement of masses and spins of new particles is straight forward.
As an example we will discuss the graviton case\cite{Allanach:2000nr}.
The graviton, which should be a spin 2 particle, can be produced
directly in proton-proton collisions at the LHC. The decay channel 
$G\to\Pelectron\Ppositron$ can be cleanly selected. The mass of
the graviton resonance can be directly measured from the di-electron
invariant mass distribution. For a given luminosity of
\unit[100]{\invfb} the graviton with $m_G$ up to \unit[2080]{GeV} can
be discovered.
\thetastar, the angle between the electron and the beam axis,
can be used to measure the spin of
the observed resonance. The general form of the $\cos\thetastar$
distribution is $1+A\cos^2\thetastar+B\cos^4\thetastar$. For graviton
production via gluons or quarks the factors are $A=0, B=-1$ and $A=-3,
B=4$, respectively. Further, the SM background is only flat ($A=0,
B=0$) for electron-pair production via a scalar resonance. In the
case of a vector resonance $A=\alpha, B=0$, where
$\alpha=1$ in the SM.
For a given luminosity of \unit[100]{\invfb} the spin 2 nature of the
graviton can be determined at 90\% CL up to graviton masses of
\unit[1720]{GeV}, which also means that the spin 1 case is ruled out.

\section{Summary}

Provided new particles are in the sub-TeV regime, already first LHC
data will allow to perform a rough spectroscopy of these. In the case
of no missing energy due to invisible particles at the end of a decay
chain, the experimental methods for mass and spin measurements are
very well established and can be applied at the LHC.
In the case of missing energy the experimental methods to measure
mass and spin of the new particles are quite advanced and will be
needed to distinguish for example SUSY from UED. Clearly, some of
the more difficult measurements need high luminosity.

%------------------------------------------------------------------------------
%       Bibliography
%------------------------------------------------------------------------------
\begin{footnotesize}
\bibliographystyle{ismd08} 
{\raggedright
\bibliography{ismd08_efeld}
}
\end{footnotesize}
\end{document}